# Comment on "Resolving spatial and energetic distributions of trap states in metal halide perovskite solar cells"

Sandheep Ravishankar,[1] Thomas Unold[2] and Thomas Kirchartz[1,3]*

[1]IEK-5 Photovoltaik, Forschungszentrum Jülich, 52425 Jülich, Germany
[2]Department of Structure and Dynamics of Energy Materials, Helmholtz-Zentrum-Berlin, Hahn-Meitner-Platz 1, 14109 Berlin, Germany
[3]Faculty of Engineering and CENIDE, University of Duisburg-Essen, Carl-Benz-Str. 199, 47057 Duisburg, Germany
*author for correspondence, email: t.kirchartz@fz-juelich.de

**Abstract**
Ni et al. (Reports, 20 March 2020, p. 1352) report bulk trap densities of $10^{11}$ cm$^{-3}$ and one to four orders change in interfacial trap densities from drive-level capacitance profiling of lead halide perovskites. From electrostatic arguments, we show that the results are not trap densities but a consequence of the geometrical capacitance and charge injection into the perovskite layer.

Despite the excellent optoelectronic properties of lead-halide perovskites, efforts to better understand the details of the remaining losses due to non-radiative recombination via defects are crucial to further improve the performance of photovoltaic or light emitting devices. One method that can determine the energetic depth of a trap and its spatial position is the so-called drive-level capacitance profiling (DLCP) method. Ni *et al*. (*1*) recently applied this method to halide perovskite solar cells to resolve bulk trap densities as low as ~$10^{11}$ cm$^{-3}$ and interfacial trap densities that increase by 1-4 orders of magnitude from bulk values (see Fig. 3A in (*1*)). However, a charge density can only be detected in capacitance measurements if it affects the electrostatic potential, which requires either sufficiently high charge densities, low permittivities or sufficient thicknesses (*2*). Using basic electrostatic arguments, we show that capacitance-based methods cannot resolve the charge densities observed in (*1*), except for the measurement shown in Fig. 1E. We show by numerical simulation that perovskite solar cells without any defects or dopant atoms yield a response that closely resembles the one in (*1*), indicating a universal threshold value below which the response cannot be considered to originate from a density of defects or dopants.

The inherent assumption required to obtain spatial information in capacitance profiling methods such as capacitance-voltage (CV) and DLCP measurements is the existence of a space-charge region of width $w$ generated by a charge density $N_\text{d}$ (dopant or trap densities) within the device of thickness $d$, that can be modified by the applied voltage $V$. Upon applying a perturbation, a response is obtained from the edge of the depletion region or from a density of emission-limited traps located at the junction transition region (*3*). Although DLCP is not a small perturbation technique like a CV measurement, the electrostatic origin of the response is identical. Indeed, the two techniques often yield similar results, especially at low frequencies where the deep traps respond (*3*).

We use this property to illustrate the limitations of the DLCP technique to resolve charge densities, from numerical simulations of CV measurements of perovskite solar cells using SCAPS (*4*). A common representation is the doping density profile, which is a plot of $N_\text{d}(w) = -2(dC^{-2}/dV)^{-1}/q\varepsilon_\text{r}\varepsilon_0$ versus profiling distance $w = \varepsilon_\text{r}\varepsilon_0/C(V)$, where $C$ is the capacitance per unit area (Fcm$^{-2}$), $\varepsilon_\text{r}$ and $\varepsilon_0$ are the relative permittivity of the perovskite and permittivity of free space respectively and $q$ is the elementary charge. A simulated doping profile for a dopant and trap free perovskite solar cell (Fig 1A; parameters and band diagram in Table S1 and Fig. S1) is shown for the same thicknesses used in Fig. 3A of (*1*). The apparent doping profile is 'U'-shaped and is nearly identical to the spatial trap density profile reported in Fig. 3A of (*1*). A similar effect is observed in Fig. 1B for an intrinsic, dopant and trap-free thin film solar cell, although the apparent doping densities are a few orders of magnitude higher, again in agreement with the values reported in (*1*).

These doping profiles can be understood from the relation between a Mott-Schottky plot ($C^{-2}$ versus $V$) and a doping profile (Fig. S2). The rise in the apparent dopant densities at the interfaces are simply the plateaus at low and high forward bias of the Mott-Schottky plots (see Fig. S3), while the apparent doping density in the bulk corresponds to the linear apparent Mott-Schottky regime. Such a shape of the Mott-Schottky profile is actually a fundamental response caused by the combination of a geometrical electrode capacitance combined with charge injection. Charge injection at forward bias in a diode typically leads to an exponentially voltage-dependent capacitance (see section A1 in the supplementary materials for further details). If we connect this capacitance in parallel to a geometric capacitance (i.e. $C = C_\text{g} + C_0 \exp(qV/mk_\text{B}T)$, where $k_\text{B}T$ is the thermal voltage and $m$ is a factor that controls the slope of $C$ vs. $V$), the shape of the doping profiles can be analytically calculated (see sections A3 and A4 in the supplementary materials).

If the doping and trap densities are too small to affect the electric field of the perovskite layer of thickness $d$, the condition $w \leq d$ is not satisfied. For example, for the lowest reported bulk trap densities of $\sim 10^{11}$ cm$^{-3}$ in $\sim 39$ µm thick perovskite layers in (*1*), the theoretical space-charge layer width at the onset of the linear Mott-Schottky region would be $w = 88.5$ µm, i.e. larger than the crystal thickness. In such situations, the geometric and injection capacitances dominate the response and yield a minimum charge density (derived in section A4 in supplementary materials) given by

$$N_\text{d,min} = 27 m k_\text{B} T \varepsilon_\text{r} \varepsilon_0 / 4 q^2 d^2. \tag{1}$$

This value (shown in Fig. 1C) sets the plateau region of the doping profile, and only measured charge densities greater than this limit (green shaded region) can be considered as a response from doping or from charged defects. Note that the condition $N_\text{d} \gg N_\text{d,min}$ holds for any measurement frequency (see also section A5 in the supplementary material). If the probed carrier or trap density does not comply with $N_\text{d} \gg N_\text{d,min}$, the capacitance response must arise from charge injection likely combined with a capacitive response of the transport layers (see section A2 in the supplementary material). Because the minimum charge density is inversely proportional to the square of the thickness of the device, intrinsic thin films will always show larger apparent doping and trap densities than bulk single crystal films, as was observed experimentally in Fig. S10B of (*1*).

As mentioned above, the apparent rise in interfacial charge densities is a direct consequence of charge injection, which is analytically described by $N_\text{d}(w) = m k_\text{B} T \varepsilon_\text{r} \varepsilon_0 / q^2 w^2$ (see section A3 in the supplementary materials) and is indeed universally observed for DLCP or CV measurements of several photovoltaic technologies (Fig. 2). Therefore, only charge densities in the plateau region of the doping profile (such as seen in Fig. 1E of (*1*)) should be considered as representing dopant/trapped charge densities if they are larger than $N_\text{d,min}$.

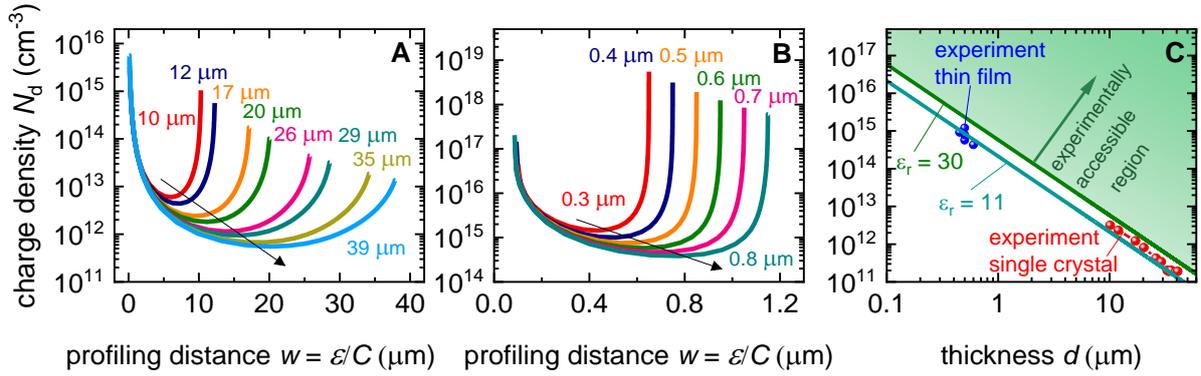

**Fig. 1. Doping profiles and minimum charge densities required for resolution in bulk single crystal and polycrystalline thin film trap-free, dopant-free perovskite solar cells.** Simulated spatial doping profiles at $10^3$ Hz of a p-i-n type PTAA (10 nm)/perovskite/PCBM (25 nm) solar cell for (**A**) same thicknesses as used in Fig. 3A in (*1*) of the bulk perovskite layer. Arrow indicates reduction of apparent bulk charge (dopant or trap) densities with increasing thickness. The profile is identical to Fig. 3A in (*1*) even with the absence of any dopant or trap densities in the model. (**B**) Different thicknesses between 300 and 800 nm representative of perovskite thin films. Arrow indicates apparent reduction of bulk charge densities with thickness. (**C**) Minimum charge densities (dopant or trap) that will be observed in a capacitance-voltage measurement ($m = 2$ is assumed) for different thicknesses and permittivities typical of perovskite (olive) and silicon or organic (cyan) solar cells, in comparison with measured minimum charge densities reported for bulk single crystal and polycrystalline thin films in (*1*). The green region represents charge densities that are experimentally accessible for the perovskite solar cell.

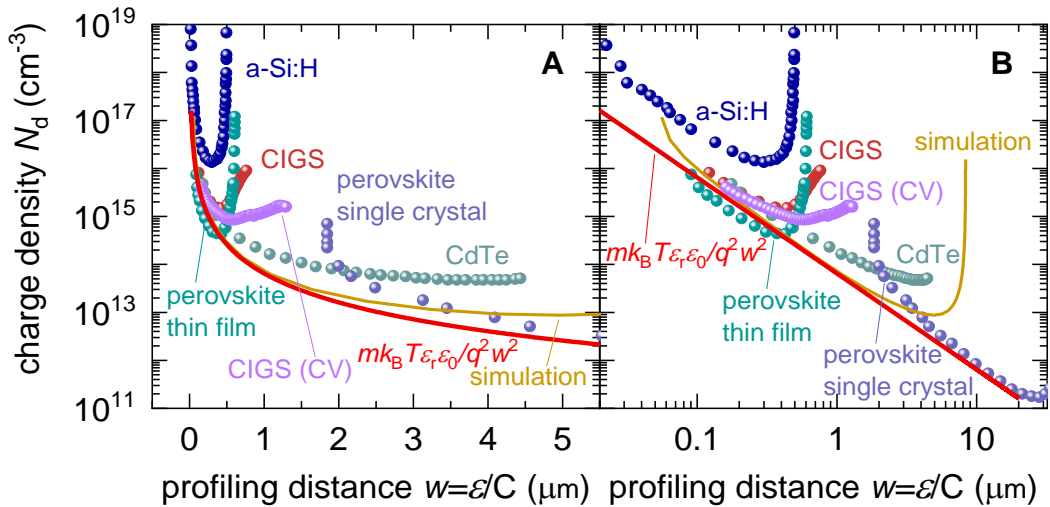

**Fig. 2. Universal rise in apparent interfacial charge densities due to charge injection.** Some reported spatial trap profiles shown with (**A**) linear and (**B**) logarithmic horizontal axis obtained from DLCP and CV measurements for different solar cell technologies such as amorphous hydrogenated silicon (a-Si:H) (*5*), copper indium gallium selenide (CuIn$_x$Ga$_{1-x}$Se$_2$ - CIGS) (*6*, *7*), methylammonium lead iodide perovskite (*1*)), cadmium telluride (CdTe) (*8*), and an 8 μm thick p-i-n (doping and trap-free) perovskite solar cell simulated using SCAPS. Also plotted is the analytical formula (with $m = 1.5$ and $\varepsilon_r = 30$) derived by considering a geometric capacitance in parallel with an exponential injection capacitance (see section A1 in the supplementary materials). The capacitance related to injection of charge at forward bias causes an apparent rise in the interfacial charge densities at the lowest profiling distances (left side of 'U'-shaped profile in A) that can erroneously be interpreted as trap densities. The geometric capacitance gives the corresponding rise in interfacial charge densities at maximum profiling distances (right side of 'U'-shaped profile in A). The universality in the doping profiles of different types of solar cells at forward bias (B) arises from the injection capacitance.

**References and notes**

1. Z. Ni *et al.*, Resolving Spatial and Energetic Distributions of Trap States in Metal Halide Perovskite Solar Cells. *Science* **367**, 1352-1358 (2020).

2. T. Kirchartz *et al.*, Sensitivity of the Mott–Schottky Analysis in Organic Solar Cells. *J. Phys. Chem. C* **116**, 7672-7680 (2012).

3. J. T. Heath, J. D. Cohen, W. N. Shafarman, Bulk and Metastable Defects in $CuIn_{1-x}Ga_xSe_2$ Thin Films using Drive-Level Capacitance Profiling. *J. Appl. Phys.* **95**, 1000-1010 (2004).

4. M. Burgelman, P. Nollet, S. Degrave, Modelling Polycrystalline Semiconductor Solar Cells. *Thin solid films* **361**, 527-532 (2000).

5. R. S. Crandall, J. Yang, S. Guha, Defect Density Profiling in Light-Soaked and Annealed Hydrogenated Amorphous Silicon Solar Cells. *Mater. Res. Soc. Symp. Proc.* **664**, A19.2 (2001).

6. L. M. Mansfield *et al.*, Comparison of CIGS Solar Cells Made with Different Structures and Fabrication Techniques. *IEEE J. Photovolt.* **7**, 286-293 (2016).

7. T. Eisenbarth, T. Unold, R. Caballero, C. A. Kaufmann, H.-W. Schock, Interpretation of Admittance, Capacitance-Voltage, and Current-Voltage Signatures in $Cu(In, Ga)Se_2$ Thin Film Solar Cells. *J. Appl. Phys.* **107**, 034509 (2010).

8. D. Menossi, E. Artegiani, A. Salavei, S. Di Mare, A. Romeo, Study of $MgCl_2$ Activation Treatment on the Defects of CdTe Solar Cells by Capacitance-Voltage, Drive Level Capacitance Profiling and Admittance Spectroscopy Techniques. *Thin Solid Films* **633**, 97-100 (2017).

9. Z. Liu *et al.*, Open-Circuit Voltages Exceeding 1.26 V in Planar Methylammonium Lead Iodide Perovskite Solar Cells. *ACS Energy Lett.* **4**, 110-117 (2018).

10. J. Maibach, E. Mankel, T. Mayer, W. Jaegermann, The Band Energy Diagram of PCBM–DH6T Bulk Heterojunction Solar Cells: Synchrotron-Induced Photoelectron Spectroscopy on Solution Processed DH6T: PCBM Blends and in Situ Prepared PCBM/DH6T Interfaces. *J. Mater. Chem. C* **1**, 7635-7642 (2013).

11. F. Staub *et al.*, Beyond Bulk Lifetimes: Insights into Lead Halide Perovskite Films from Time-Resolved Photoluminescence. *Phys. Rev. Appl.* **6**, 044017 (2016).

12. L. M. Herz, Charge-Carrier Mobilities in Metal Halide Perovskites: Fundamental Mechanisms and Limits. *ACS Energy Lett.* **2**, 1539-1548 (2017).

13. S. M. Sze, K. K. Ng, *Physics of Semiconductor Devices* (John Wiley & sons, 2006),

pp.100-102.

14. SCAPS definition files of all simulations in this Comment are deposited in the Zenodo archive with the following identifier: 10.5281/zenodo.4116604.

## Acknowledgements

**Funding:** Sandheep Ravishankar and Thomas Kirchartz acknowledge funding from the Helmholtz Association via the project PEROSEED. Thomas Unold acknowledges the DFG-SPP-2169 Peroint project for funding.

**Author contributions:** Sandheep Ravishankar performed the SCAPS simulations, made the figures and wrote the major part of the text. Thomas Unold and Thomas Kirchartz contributed to writing the text and deriving equation 1. All authors discussed the text and the data.

**Competing interests:** The authors have no competing interests to declare.

**Data and materials availability:** All data are available in the manuscript or in the supplementary material. SCAPS definition files of all simulations in this Comment are archived in Zenodo with the following identifier: 10.5281/zenodo.4116604.

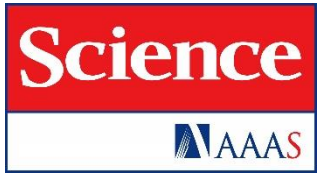

Supplementary Materials for

# Comment on "Resolving Spatial and Energetic Distributions of Trap States in Metal Halide Perovskite Solar Cells"

*Sandheep Ravishankar, Thomas Unold and Thomas Kirchartz*

correspondence to: t.kirchartz@fz-juelich.de

**This PDF file includes:**

Materials and Methods
Figs. S1 to S6
Section A1 to A5
Table S1
Discussion of the parameters

**Materials and Methods**

<u>Methods</u>

All the simulations were carried out using the program SCAPS (A Solar Cell Capacitance Simulator).

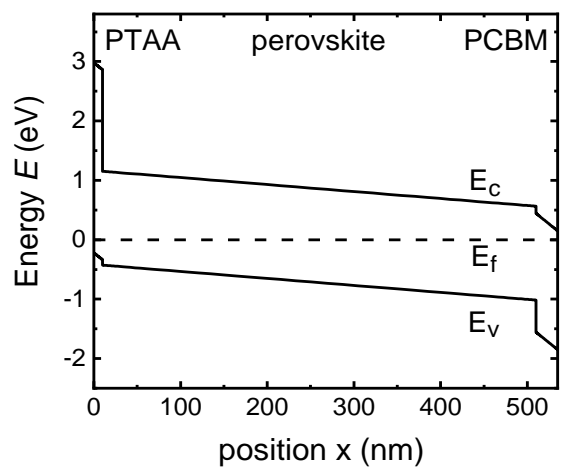

**Fig. S1.**
**Reference band diagram.** Band diagram of a reference PTAA (10 nm)/CH$_3$NH$_3$PbI$_3$ perovskite (500 nm)/ PCBM (25 nm) stack.

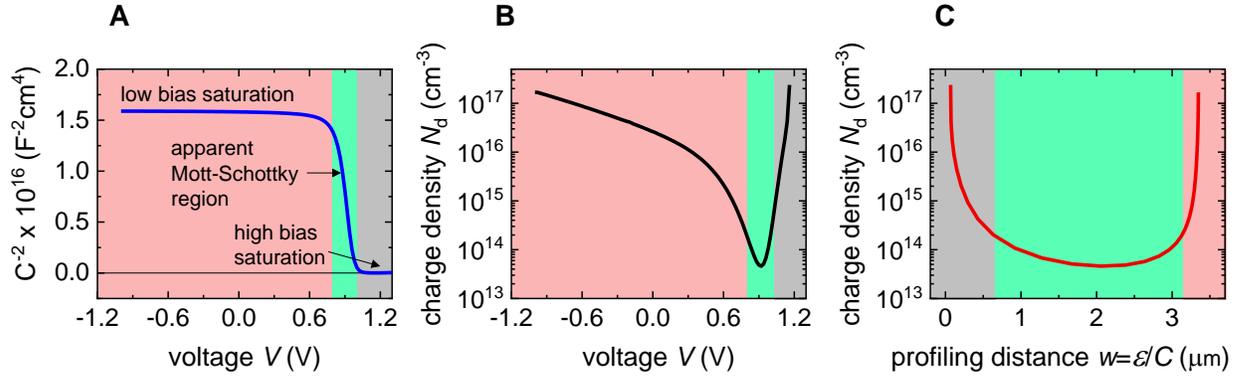

**Fig. S2**
**Connection between the Mott-Schottky plot and doping profile plot.** (**A**) Simulated Mott-Schottky plot of a p-i-n type PTAA (10 nm)/ perovskite (3 µm)/ PCBM (25 nm) solar cell with its corresponding apparent doping profile versus voltage shown in (**B**) and versus profiling distance shown in (**C**). The Mott-Schottky plot is linear at intermediate voltages while saturating to an almost constant value at forward and reverse bias as shown. The doping profile is proportional to the inverse slope of the Mott-Schottky plot, given by $N_\mathrm{d}(V) = -2(dC^{-2}/dV)^{-1}/q\varepsilon_\mathrm{r}\varepsilon_0$. This is shown as a function of voltage in **B**. However, a common representation to obtain spatial information is plotting the x-axis as the profiling distance $w$ given by $w = \varepsilon_\mathrm{r}\varepsilon_0/C(V)$. This is shown in **C**, where peaks are formed in the saturation regimes of **A** corresponding to apparent maximum and minimum profiling distances (i.e.: interfaces of the perovskite) and a plateau-like evolution in the apparent Mott-Schottky region in the bulk.

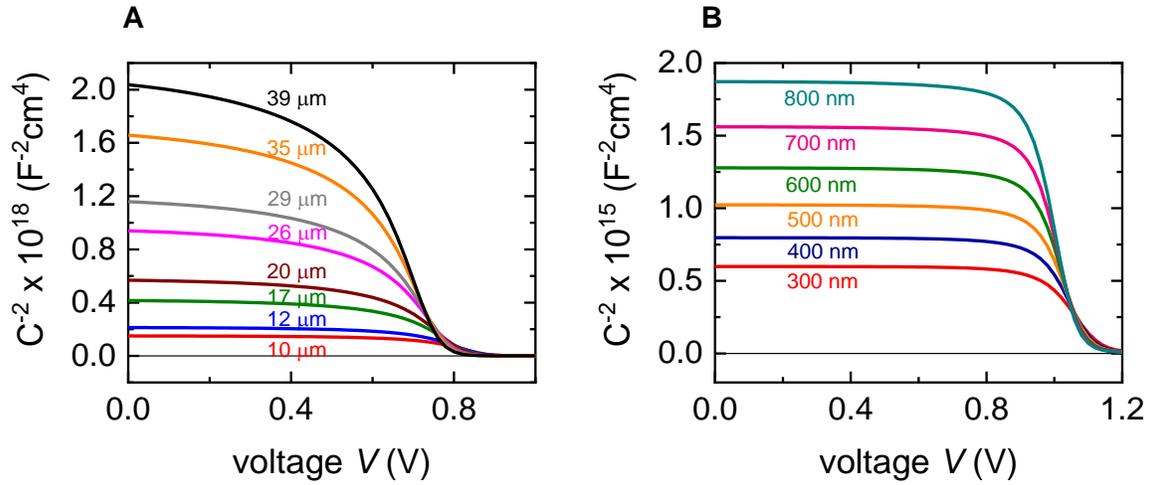

**Fig. S3**
**Mott-Schottky plots of bulk and thin film trap-free, dopant-free perovskite solar cells.** Simulated Mott-Schottky plots at $10^3$ Hz of a p-i-n type perovskite solar cell with no doping or traps for (**A**) varying thicknesses of the bulk perovskite layer and (**B**) thin film perovskite layer. These plots correspond to the capacitance-voltage simulations of Fig. 1A, B in the main text respectively.

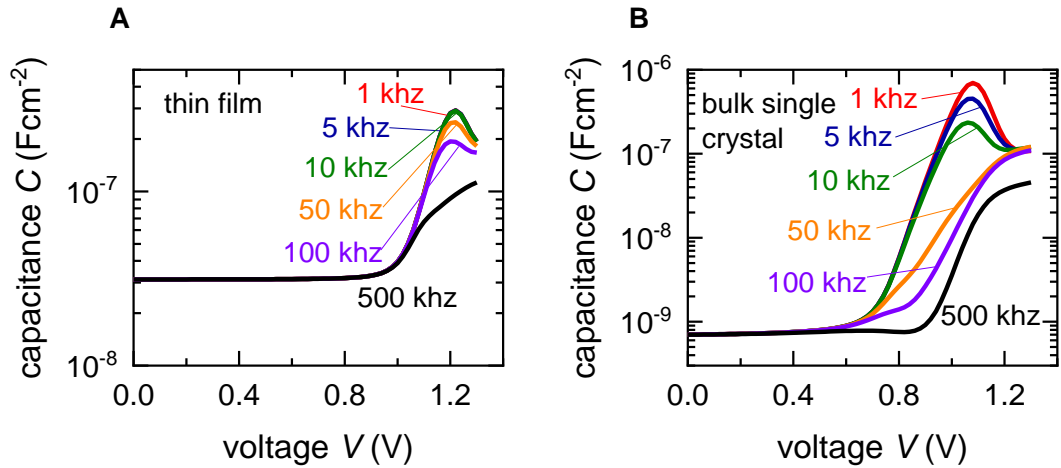

**Fig. S4**

**Frequency dependence of capacitance-voltage plots of dopant-free, trap-free thin film and bulk single crystal perovskite solar cells.** Simulated capacitance-voltage plots at different frequencies of (**A**) thin film (500 nm thick perovskite layer) and (**B**) bulk single crystal (39 µm thick perovskite layer) perovskite solar cells with no doping or traps. A frequency dependent capacitance is observed only at large applied forward biases.

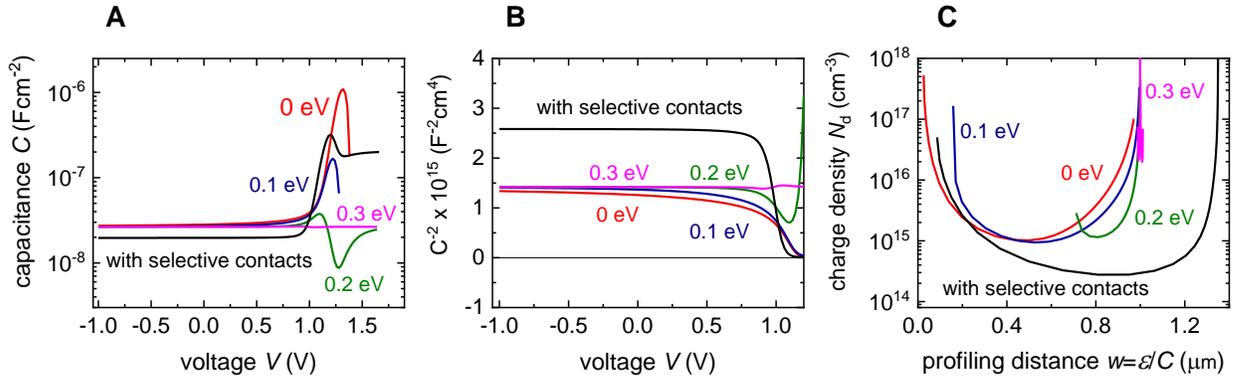

**Fig. S5**
**Effect of chemical capacitance on doping profile.** Simulated (**A**) capacitance-voltage plots (**B**) corresponding Mott-Schottky plots and (**C**) doping density profiles of a p-i-n type PTAA (10 nm)/ perovskite (1 µm)/ PCBM (25 nm) solar cell with and without electron and hole selective contact layers (PCBM and PTAA respectively) and without any doping or traps in the perovskite layer. For the case without selective contact layers, the injection barrier for electrons and holes at cathode and anode respectively are set equal and varied as shown. Smaller injection barriers show a larger contribution of the chemical capacitance due to injection of carriers at large forward bias. For large injection barriers beyond ∼ 200 meV, the chemical capacitance is not seen and the response is solely from the constant geometric capacitance measured at reverse bias. This is reflected in the doping density profiles in C where the apparent profiling distance is almost a constant value for the 0.3 eV barrier case but gradually scans a bigger range of apparent profiling distances for smaller barriers. Note also that the capacitance with selective contact layers is different in magnitude to that without selective contact layers, which affects the apparent profiling distance.

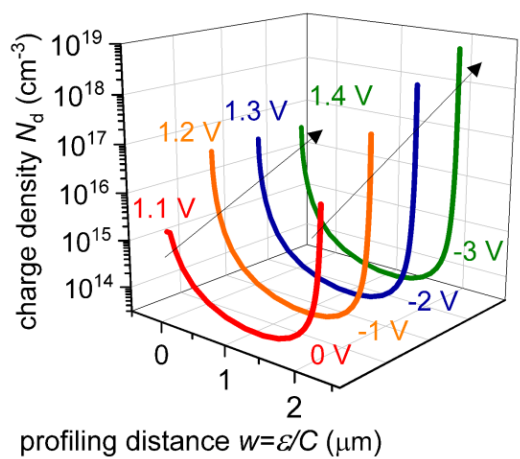

**Fig. S6**
**Sensitivity of apparent interfacial charge densities to applied voltage in doping profiles.** Evolution of apparent interfacial charge densities (sharp upward peaks) for a p-i-n type PTAA (10 nm)/perovskite (2 µm) /PCBM (25 nm) solar cell due to maximum forward and reverse bias voltage. Increments of voltage at deep reverse bias give a sharp upward rise in the apparent charge densities due to the constant geometric capacitance, while forward bias voltage increments yield larger apparent charge densities due to the chemical capacitance, further discussed in section A3.

## Section A1.

### Voltage-Dependent Capacitances in Solar Cells

At forward bias in a diode, charge is injected into the active layers of the diode and will at some point influence the capacitance. Just like the forward injection current is exponential with voltage, also the capacitance will in most cases be exponential with voltage for a certain voltage range. However, the terminology, the analytical equations as well as the exact mechanism for capacitance changes due to charge injection at forward bias differ depending on the type of diode or solar cell in question. Given that there are readers with different backgrounds, we will just briefly introduce the three most common effects and terms to help understand, why we use a voltage dependent capacitance to derive equation 1 in the main paper. These three types of capacitances are the diffusion capacitance known from the theory of pn-junction diodes, the chemical capacitance which is a similar concept used in the absence of a large neutral zone and finally, we will discuss capacitances that result from the coexistence of different layers in the solar cell that have voltage dependent resistances. Note that we will give analytical equations or discuss simple models to better explain the features and the behaviour of different capacitance types. However, given that the equations and models are often too simplistic to directly represent the superposition of various effects often seen in devices, we always use the numerical drift-diffusion solver SCAPS for illustration of the principles.

### Diffusion capacitance

A fundamental capacitance that originates from charge injection and diffusion in a p-n junction is the diffusion capacitance (*13*). The derivation of the diffusion capacitance requires solving the diffusion equation with complex quantities (including complex charge carrier lifetimes) and is valid in the base (neutral zone) of a pn-junction. The classical example is a crystalline Si solar cell (or diode), where the neutral zone is typically three orders of magnitude thicker than the space charge region. The low frequency limit is given by

$$C_{\text{diff}} = \frac{q}{k_\text{B}T}\left(\frac{qD_\text{p}p_{\text{n}0}}{L_\text{p}} + \frac{qD_\text{n}n_{\text{p}0}}{L_\text{n}}\right)\exp\left(\frac{qV}{mk_\text{B}T}\right) = C_0 \exp\left(\frac{qV}{mk_\text{B}T}\right), \quad (\text{S1})$$

where $D_\text{p}$, $D_\text{n}$ and $L_\text{p}$, $L_\text{n}$ are the diffusion coefficients and diffusion lengths respectively of holes and electrons and $n_{\text{p}0}$, $p_{\text{n}0}$ are the equilibrium minority carriers concentrations in the p and n sides of the p-n junction respectively. While the diffusion capacitance already provides a simple analytical explanation for the exponential voltage dependence of capacitances at forward bias, the absolute value of the diffusion capacitance requires the presence of a doped semiconductor layer with a clearly defined minority carrier concentration. Therefore, the absolute values predicted by equation S1 are not applicable to devices with mainly intrinsic or lowly doped semiconductor layers such as perovskite solar cells.

### Chemical capacitance

In case of a doped semiconductor layer of thickness $d$ between two contacts, a closely related quantity to the diffusion capacitance is the chemical capacitance, related to injection of minority carriers $n$ into the semiconductor. The chemical capacitance is often used in the context of organic solar cells with large built-in electric fields that separate electrons from holes. In this case, the concentrations of electrons and holes often change exponentially in a certain part of the device and for a certain voltage range, which then leads to an exponential change of capacitance via

$$n = n_0 \exp(\frac{qV}{mk_\text{B}T}), \tag{S2}$$

$$C_\mu = qd\frac{dn}{dV} = \frac{q^2 nd}{mk_\text{B}T} = C_0\exp(\frac{qV}{mk_\text{B}T}). \tag{S3}$$

This capacitance is added to the geometric capacitance of the device to form the general capacitance-voltage relation

$$C(V) = C_\text{g} + C_0\exp(\frac{qV}{mk_\text{B}T}). \tag{S4}$$

Capacitances in multi-layer systems

Most solar cells have more than one semiconductor layer that is sufficiently resistive that its capacitance may contribute to the capacitance of the total device. In a perovskite solar cell of the design seen in the main paper, we have the hole transport layer (HTL) PTAA, the perovskite and the fullerene-based electron transport layer (ETL) as three layers that each have a capacitance and a differential resistance at each working point. These three RC circuits are connected in series and the total capacitance results from this series connection of three RCs. It can be easily shown that the imaginary part of the admittance of three series connected RCs depends also on the resistances and not only on the capacitances of the three RC circuits (see equations S5 and S6). This can be rationalized by imagining one of the three resistances to go to zero. In this case, also the capacitance of this RC would not matter anymore, because all the current could flow via the short circuit formed by the 0 Ohm resistance. This leads to the simple consequence that voltage dependent resistances may cause voltage dependent capacitances in a multilayer system. In addition, even if none of the capacitances was voltage dependent, the corresponding resistances of each layer will certainly be voltage dependent. In particular, the differential resistance of the perovskite layer is its recombination resistance that varies exponentially with voltage just as the recombination current does. Hence, at low forward bias, the resistance of the perovskite layer is huge, while at high forward bias the resistance will become tiny, thereby essentially short circuiting the RC of the perovskite. This will not happen to the same degree to the RCs of the ETL and HTL. Their recombination resistances will also vary with voltage but to a lesser degree and at large forward bias, the capacitance will most likely be dominated by either or both of the capacitances of the ETL and HTL. Thus, charge injection and subsequent recombination in a multilayer stack may lead to an exponentially voltage dependent capacitance (in a certain range of voltages) because the differential resistances are exponentially voltage dependent even if the capacitances were mainly geometrical. If the capacitances have additional exponential contributions because of an exponential increase in charge-carrier density, the two effects will overlap but generally still result in an exponential voltage dependence of the total capacitance at forward bias. The superposition of these effects makes it extremely difficult to assert with certainty which exact mechanisms are causing the voltage dependence of the forward bias capacitance in most perovskite solar cells. At the same time, all of them will have an exponential voltage dependence, which therefore simplifies the mathematical description of the phenomenon.

In Fig. S5, we compare the effect of having several layers vs. a single layer in numerical simulations with SCAPS (assuming no doping or traps in the perovskite layer). Note that these simulations do not use any of the analytical equations discussed above but solve the full continuity and Poisson equations. The black line shows a simulation as done in the rest of the paper and supplementary material that includes thin PTAA and PCBM layers in addition to a perovskite absorber layer (with selective contacts). This data shows first a relatively constant capacitance (geometrical capacitance of the three layers in series) that then increases strongly at around 0.8 V leading to the typical U-shape for the charge density profile. This trend is at least partly due to the voltage dependent recombination resistance of the perovskite layer. If we omit the two charge selective layers, we only have a single semiconductor layer with charge injection barriers for electrons and holes on opposite sides. Here, we still observe a strongly voltage dependent capacitance and the typical U-shape in panel C, but the result now depends strongly on the injection barrier. The lower the barrier, the higher the concentration at the charge injecting contact and the higher the capacitance will rise. This is consistent with the general behaviour of chemical capacitances in metal-semiconductor-metal type diodes, where the voltage dependence of the capacitance is related to the voltage dependent injection of charge carriers. Note also that for the cell with selective contacts, the profiling axis extends beyond the total thickness of the device. This is a consequence of the key assumptions for deriving the profiling depth not holding anymore. In particular, there is no longer one single permittivity but different permittivities for different layers.

Another consequence of the injection capacitance in multilayer systems is that there is always a frequency dependence of the total capacitance even if the individual elements making up the equivalent circuit are not frequency dependent. Thus, if we just look at the simplified model of three RC-circuits in series (where each RC consists of a resistance and a capacitor in parallel), the resulting $C$ of the total circuit will be frequency dependent if at least one of the resistances is voltage dependent. This is due to the fact that the imaginary parts $\omega C$ of the admittance of each RC circuit are automatically frequency dependent, via the frequency dependence of $\omega$. While this frequency dependence is not a result of trap states, it may of course affect the interpretation of all frequency dependent capacitance measurements.

The admittance $Y$ of this model is given by

$$Y(V, \omega) = [(\frac{1}{R_{\text{ETL}}(V)} + i\omega C_{\text{g,ETL}})^{-1} + (\frac{1}{R_{\text{PVK}}(V)} + i\omega C_{\text{g,PVK}})^{-1} + (\frac{1}{R_{\text{HTL}}(V)} + i\omega C_{\text{g,HTL}})^{-1}]^{-1}, \quad \text{(S5)}$$

where $R$ and $C_{\text{g}}$ are the resistance and geometric capacitance respectively. This yields the capacitance as

$$C(V, \omega) = \frac{Im(Y)}{\omega} = Im[\frac{1}{\omega}(\frac{R_{\text{ETL}}(V)}{1+i\omega R_{\text{ETL}}(V)C_{\text{g,ETL}}} + \frac{R_{\text{PVK}}(V)}{1+i\omega R_{\text{PVK}}(V)C_{\text{g,PVK}}} + \frac{R_{\text{HTL}}(V)}{1+i\omega R_{\text{HTL}}(V)C_{\text{g,HTL}}})^{-1}] \quad \text{(S6)}$$

**Section A2.**

**Capacitance-voltage vs. DLCP – a brief comparison**

Drive level capacitance profiling (DLCP) uses a variation of the AC voltage to generate additional information relative to the classical capacitance voltage profiling method, which is based on a small signal analysis. The DLCP method uses two terms in the Taylor expansion of capacitance vs. AC voltage amplitude $\Delta V$, namely the zero and first order term, such that the capacitance at each frequency and DC voltage can be written as $C = C_0 + C_1 \Delta V$, while the capacitance voltage measurement would use $C = C_0$. Since $C_1$ can have a slightly different frequency dependence than $C_0$, the DCLP data and the capacitance-voltage data may differ.

We note that in the paper by Ni et al., the DLCP data (e.g. Fig. 2A) shows a small frequency dependence throughout the whole voltage range, while the capacitance-voltage data (see Fig. S3) shows a negligible frequency dependence at small forward bias and slightly stronger frequency dependence at higher forward bias. The same can also be seen in our numerical simulations shown in Fig. S4. Ni et al. interpret the high frequency data as originating from free carriers and the difference between low and high frequency data as originating from trapped charge. In both cases, the spatial dependence would be basically identical, showing the characteristic U-shape seen and discussed in Fig. 1 of our comment, with the densities rising by orders of magnitude towards the two contacts. In our opinion, it is not credible that the U-shape seen at all frequencies is an actual consequence of the doping density (and consequently the free carrier density) as well as the trap density having the exact same shape. Instead, it is all a consequence of the combination of geometrical and exponentially voltage dependent capacitances due to charge injection as discussed quantitively in sections A3 and A4.

## Section A3.

**Analytical expression for doping profile at forward bias**

We consider a general capacitance of the form (*13*)

$$C = C_g + C_0 \exp\left(\frac{qV}{mk_BT}\right). \tag{S7}$$

where $C_g = \varepsilon_r\varepsilon_0/d$ is the geometric capacitance of the layer of thickness $d$ and $C_0$ and $m$ are the pre-factor and factor that controls the slope of the capacitance versus voltage respectively, for the diffusion capacitance that is considered proportional to the current-voltage (injection) characteristics of the diode. If the formalism developed for reverse bias capacitance is applied to this capacitance, the apparent doping density profile is given by

$$N_d = \frac{-2}{q\varepsilon_r\varepsilon_0}\left[\frac{dC^{-2}}{dV}\right]^{-1}, \tag{S8}$$

which can be represented in terms of the slope of the capacitance versus voltage and profiling position $w = \varepsilon_r\varepsilon_0/C$ as

$$N_d = \frac{\varepsilon_r\varepsilon_0 C}{qw^2}\left[\frac{dC}{dV}\right]^{-1}. \tag{S9}$$

For large forward bias, we have

$$C \cong C_0 \exp\left(\frac{qV}{mk_BT}\right), \tag{S10}$$

$$\frac{dC}{dV} = \frac{qC}{mk_BT}. \tag{S11}$$

Substituting equations S10 and S11 in equation S9, we get

$$N_d(w) = \frac{mk_BT\varepsilon_r\varepsilon_0}{q^2w^2}. \tag{S12}$$

Equation S12 shows that $N_d \propto w^{-2}$ at large forward bias, which explains the rise in interfacial charge densities for the lowest profiling distances (i.e. close to the interface). This forms the left side of the 'U'-shaped doping profile. The flat region in the profile can be described by a constant value $N_{d,min}$, while the constant geometric capacitance gives an infinite rise in charge densities at reverse bias, forming the right side of the 'U'-shaped profile. Therefore, the doping profile at forward bias is given by

$$N_d(w) = N_{d,min} + \frac{mk_BT\varepsilon_r\varepsilon_0}{q^2w^2}. \tag{S13}$$

## Section A4.

### Derivation of minimum charge density for resolution

We again consider a general capacitance

$$C = C_\text{g} + C_0 \exp(\frac{qV}{mk_\text{B}T}), \tag{S14}$$

The doping density profile is given by

$$N_\text{d} = \frac{-2}{q\varepsilon_\text{r}\varepsilon_0}[\frac{dC^{-2}}{dV}]^{-1}. \tag{S15}$$

The profiling position is given by

$$w = \frac{\varepsilon_\text{r}\varepsilon_0}{C}, \tag{S16}$$

and at deep reverse bias, we obtain the thickness of the layer as

$$d = \frac{\varepsilon_\text{r}\varepsilon_0}{C_\text{g}}. \tag{S17}$$

Substituting equations S14 and S17 in S15, we get

$$N_\text{d} = \frac{mk_\text{B}T\varepsilon_\text{r}\varepsilon_0 C^3}{q^2 d^2 C_\text{g}^2 C_0 \exp(qV/mk_\text{B}T)}. \tag{S18}$$

To obtain the minimum value of the doping density, we need to solve

$$\frac{dN_\text{d}}{dw} = \frac{dN_\text{d}}{dV}\frac{dV}{dw} = 0. \tag{S19}$$

Differentiating equation S16 and S18 with respect to voltage, we obtain

$$\frac{dV}{dw} = \frac{-C^2}{\varepsilon_\text{r}\varepsilon_0(dC/dV)}, \tag{S20}$$

$$\frac{dN_\text{d}}{dV} = \frac{mk_\text{B}T\varepsilon_\text{r}\varepsilon_0}{q^2 d^2 C_\text{g}^2 C_0}[\frac{3C^2(dC/dV)\exp(qV/mk_\text{B}T) - (qC^3/mk_\text{B}T)\exp(qV/mk_\text{B}T)}{\exp(2qV/mk_\text{B}T)}]. \tag{S21}$$

Solving equation S19 using equations S14, S20 and S21, we obtain

$$C_\text{min} = 3C_0 \exp(qV_\text{min}/mk_\text{B}T), \tag{S22}$$

which is the minimum value of the capacitance at a corresponding voltage $V_\text{min}$. Substituting equation S22 in equation S14, we obtain

$$C_\text{g} = 2C_0 \exp(qV_\text{min}/mk_\text{B}T). \tag{S23}$$

Substituting equations S22 and S23 in equation S18 at the voltage $V_\text{min}$, we obtain the minimum doping density as

$$N_\text{d,min} = \frac{27 mk_\text{B}T\varepsilon_\text{r}\varepsilon_0}{4q^2 d^2}. \tag{S24}$$

**Section A5.**

**Frequency dependence of $N_{d,min}$**

The derivation of eq. (S24) is based on the argument of a geometrical capacitance, which is basically frequency independent, and a capacitance due to charge injection. As we explain in section A1, the capacitance due to charge injection is expected to have a (moderate) frequency dependence that can be observed both in Fig. S3 of (*1*) and in our simulations shown in Fig. S4. This frequency dependence of the capacitance due to charge injection results in small changes of the prefactor *m* that controls the voltage dependence of the capacitance due to charge injection. Thus, we expect small changes of the value of $N_{d,min}$ ~ *m* with frequency. However, the prefactor *m* and subsequently $N_{d,min}$ will not change by orders of magnitude with frequency and the important trend with thickness that is so prominently observed in (*1*) is not affected.

| parameter | PTAA | perovskite | PCBM |
| --- | --- | --- | --- |
| thickness (nm) | 10 nm | variable | 25 nm |
| relative permittivity | 3 | 30 | 3 |
| bandgap (eV) | 3.2 | 1.58 | 2 |
| electron affinity (eV) | 2.22 | 3.93 | 4.05 |
| effective DOS CB (cm$^{-3}$) | $2 \times 10^{18}$ | $2 \times 10^{18}$ | $2 \times 10^{18}$ |
| effective DOS VB (cm$^{-3}$) | $2 \times 10^{18}$ | $2 \times 10^{18}$ | $2 \times 10^{18}$ |
| radiative recombination coefficient (cm$^3$/s) | 0 | $6 \times 10^{-11}$ | 0 |
| electron mobility (cm$^2$/Vs) | $10^{-3}$ | 20 | $10^{-3}$ |
| hole mobility (cm$^2$/Vs) | $10^{-3}$ | 20 | $10^{-3}$ |
| doping density (cm$^{-3}$) | 0 | 0 | 0 |

**Table S1.**

Parameters used for the simulations in the main text and supplementary materials.

**Discussion of the parameters**

   **Metal contacts**: The metal contact workfunctions chosen were 5.2 eV (ITO) for the PTAA layer side and 4.2 eV (Ag) for the PCBM layer side to obtain a built-in voltage of 1 V. The surface recombination velocities for electrons and holes at both metal contacts was set to $10^7$ cm/s.

   **Thickness**: The thicknesses were chosen from (*1*).

   **Relative permittivity**: The relative permittivity value for the PTAA and PCBM layers was set to 3 since typical values for fullerenes lie between 2 and 4.

   **Bandgap**: The PTAA layer bandgap was chosen from (*9*). The perovskite layer considered was a $CH_3NH_3PbI_3$ perovskite with a bandgap of ~1.58 eV. The PCBM layer bandgap was chosen based on ref. (*10*).

   **Electron affinities**: The PCBM layer electron affinity was set to 4.05 eV based on the different values reported (between 3.7 and 4.2 eV) and considering the Ag work function of 4.2 eV. The perovskite layer electron affinity was increased from 3.83 eV (obtained from (*9*)) to 3.93 eV to reduce the barrier for electrons at the perovskite/PCBM interface.

   **Effective density of states (DOS)**: The effective DOS for the conduction and valence band of the perovskite layer was chosen from (*11*). The DOS of the PCBM and PTAA layers were chosen to be the same as that of the perovskite layer.

   **Radiative recombination coefficient**: The order of the perovskite layer radiative recombination coefficient was chosen from (*11*). No recombination in the PCBM and PTAA layers was assumed.

   **Mobility**: We fixed the electron and hole mobilities to be equal in all cases for simplicity. Based on the generally large mobilities reported for perovskite layers, we fixed a value of 20 cm$^2$/Vs (*12*).